\documentclass{IEEEtran}
\usepackage{graphicx}

\usepackage{balance}

\begin{document}

\title{Enabling Opportunistic Users in Multi-Tenant IoT Systems using Decentralized Identifiers and Permissioned Blockchains}

\author{
  \IEEEauthorblockN{Nikos Fotiou, Iakovos Pittaras, Vasilios A. Siris, George C. Polyzos\\}
  \IEEEauthorblockA{
    Mobile Multimedia Laboratory,\\
    Department of Informatics,\\
    Athens University of Economics and Business\\
    Email:\{fotiou, pittaras, vsiris, polyzos\}@aueb.gr
    }
}

\maketitle

\begin{abstract}
In this work, we leverage advances in decentralized identifiers and permissioned blockchains to build a flexible user authentication and authorization mechanism that offers enhanced privacy, achieves fast revocation, and supports distributed ``policy decision points'' executed in mutually untrusted entities. The proposed solution can be applied in multi-tenant ``IoT hubs'' that interconnect diverse IoT silos and enable authorization of ``guest'' users, i.e., opportunistic users that have no trust relationship with the system, which has not encountered or known them before.
\end{abstract}



\maketitle

\section{Introduction}
As the Internet of Things (IoT) emerges it becomes clear that it departs from the traditional networking architectures. IoT devices are deprived from computational power, have limited connection capabilities, and in many use cases they are physically exposed (hence more vulnerable). For these reasons, in most IoT systems users do not interact directly with the IoT devices, instead their communication is mediated by a more powerful gateway.  Nevertheless, gateways and devices are usually bundled into the silo of a specific manufacturer. What is worse, these silos are vertical and not interoperable. The need for interoperability, e.g., among home IoT systems, has fostered a new type of IoT devices, usually referred to as \emph{IoT hubs}. IoT hubs, such as Amazon's echo and Google's home, allow users to interact with multiple IoT gateways through a single device (the hub) using a unified interface.

It is evident that traditional authentication and authorization systems cannot be applied in this environment. Indeed, recently, He et al.~\cite{He2018} pinpointed the need for novel access control mechanisms for the IoT that will handle  these new entities and trust relationships, taking at the same time into consideration users' context, as well as properties of the users' environment. To this end, we design and build an IoT access control solution that supports authorization of opportunistic users, referred to as guests, with rapid revocation and enables the evaluation of complex access control policies by mutually untrusted entities. In order to achieve our goal we leverage recent advances in decentralized identifiers and permissioned blockchain technologies. 

\section{System design}
\begin{figure}
\includegraphics[width=0.9\linewidth]{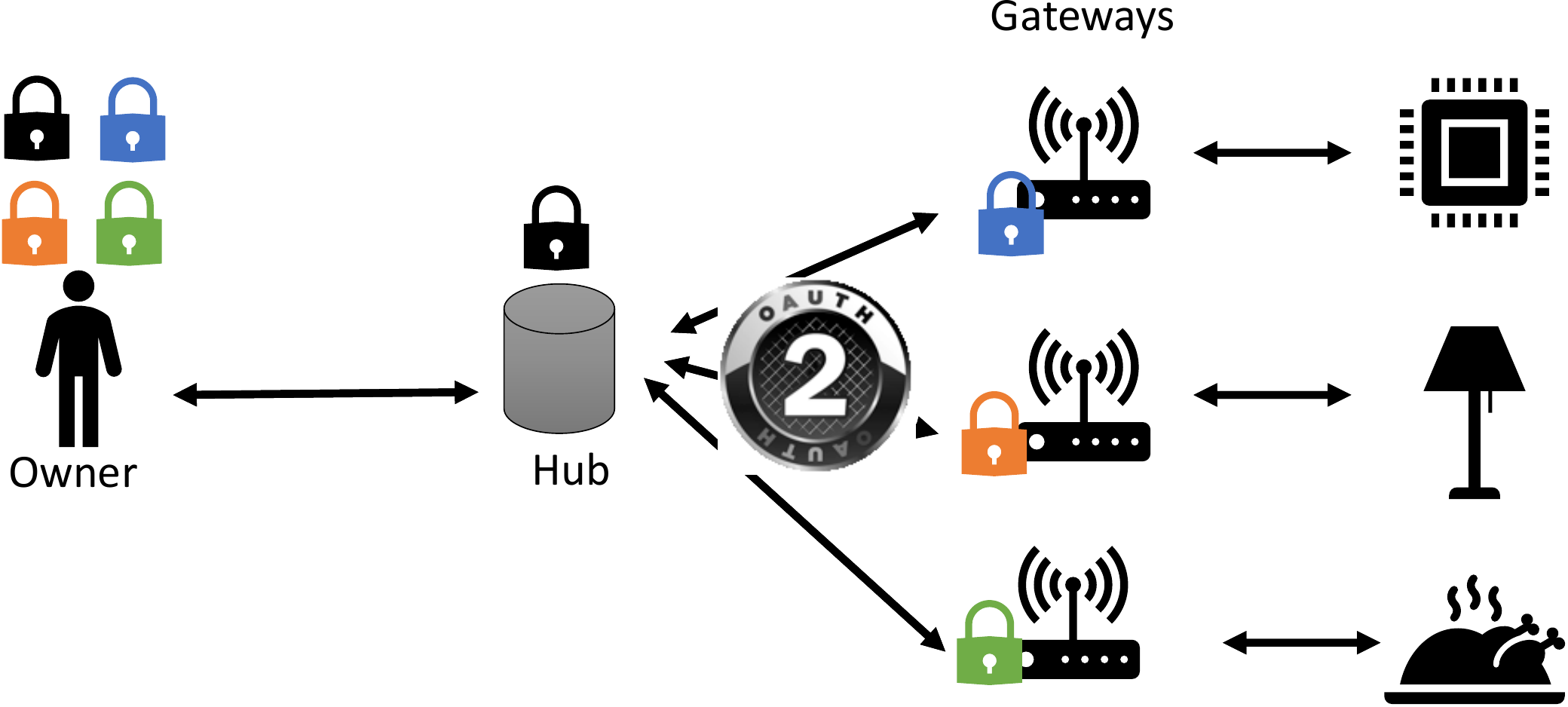}
\caption {Our reference IoT architecture}
\label{fig:arch}
\end{figure}
Our solution considers the architecture depicted in Figure~\ref{fig:arch}. In this architecture there is a user, referred to as the \emph{owner}, who interacts with IoT gateways through an IoT hub. Each gateway and the hub have their own authentication and authorization systems; hence the owner has an ``account'' in all these systems. Furthermore, the owner has ``linked'' the hub account with all other accounts using OAuth~\cite{oauth} (or any other similar protocol). In other words, the owner has authorized the IoT hub to communicate with the IoT gateways on his behalf. 
Our system achieves the following:

\begin{itemize}
\item Guests can interact with the IoT gateways through the hub.
\item Guests cannot be tracked when interacting with hubs belonging to different owners.
\item Guests' access rights can be easily modified/removed.
\item Gateways do not have to be modified.
\end{itemize}

Our system relies on Decentralized Identifiers (DIDs) for guest authentication.  A DID is a new type of identifier that is globally unique, resolvable with high availability, and cryptographically verifiable~\cite{did-primer}. From a high level perspective a DID system can be viewed as a key-value storage system, where the key is the DID and the value is a \emph{DID document}. DID documents can be stored in blockchains, distributed ledgers, (decentralized) P2P networks, or other systems with similar capabilities; these systems are referred to as \emph{Decentralized Identifier Registries} (but we simply refer to them as registries).  DID documents contain (among other information) 
pubic keys that can be used for linking a user to particular DID document (using an ``authentication'' that depends on the type of the public key), service endpoints, as well as auxiliary information that can be used for verifying the integrity of the document~\cite{did-spec}. 

\section{Simple user authorization}
In our system we assume a secure DID registry which is trusted and can be accessed at least by the owners and the hubs. Furthermore, we assume that each owner is associated with a DID ($DID_{owner}$), known to the hub, and mapped to a public key ($P_{owner}$) used for singing the DID documents of the guests. A guest wishing to access an IoT resource through a hub must initially create a public-private keypair and securely transmit the public key $P_{guest}$ to the owner. Then, the owner generates a DID ($DID_{guest}$), a DID document, signs the document and stores it in the registry, and sends back the DID to the guest. The DID document contains $P_{guest}$ and the corresponding authentication method, the URIs of the IoT resources the guest can access, as well as an expiration time.

IoT resource access (by a guest) is a two-step process: firstly, the guest is authenticated and authorized and secondly the guest invokes the appropriate hub API function. Guest authentication and authorization is implemented as follows.  Initially, the guest sends to the hub her $DID_{guest}$. Then, the hub, retrieves the corresponding DID document from the registry, checks if it has been signed by $DID_{owner}$, verifies that the document is still active, and retrieves $P_{guest}$. As a next step the hub executes the authentication method that corresponds to the type of $P_{guest}$.

\section{Authorization delegation}
Our first construction albeit it can be useful for simple use cases it cannot easily accommodate cases that require complex access control policies or cases where IoT hubs are not trusted by guests to evaluate a policy. These problems stem from the fact that in our first model the IoT hub holds both the role of the \emph{Policy Decision Point} (PDP)  and of the \emph{Policy Enforcement Point} (PEP)~\cite{rfc2753}.  With the approach we present in this section we separate these two functionalities and we assign to the IoT hub only the role of the PEP. 

In our authorization delegation solution we model access control policies as a function, which accepts arbitrary inputs (e.g., a resource URI, a $DID_{guest}$, context related information) and outputs a Boolean value, which indicates if the authorization procedures was successful, and (optionally) a timestamp that defines the validity period of the response. These functions are identified by a URI and they can be invoked using remote procedure calls; we refer to this URI as $Policy_{URI}$. Supposedly that there is a PDP trusted by both the owner and the guest, then our solution can be easily implemented using access control delegation mechanisms (e.g., similar to~\cite{Fot2016}). However, if this is not the case then the PDP can be implemented as a distributed process where all policy decisions are collected to a single point and the final access control decision is made upon consensus. In this work we consider the latter approach, i.e., we support multiple PDPs that all, at the same time, execute the same policy, and the final output is decided based on a pre-defined consensus algorithm. It should be noted that the IoT hub is oblivious about the access control decision process: from its perspective the PDP decision is received by making a single remote procedure call. 

Similar to our simple authorization construction, IoT resource access involves guest authentication, authorization, and API invocation. When authorization delegation is used, the guest authorization process is modified as follows. The first time a guest requests access to an IoT resource, the hub performs a remote procedure call to the corresponding $Policy_{URI}$  (found in the DID document), receives, and stores the result: If the result of this call is positive then the guest is considered authorized to access the requested resource.

\section{Conclusions}
We propose an access control solution for modern IoT systems. Our solution leverages decentralized identifiers and permissioned blockchains to facilitate controlled access by ``guest'' users. Our designs are tracking-resistant, they facilitate revocation, and allow fine-grained access control decisions, even by mutually untrusted entities. With our solution, guests are granted access to IoT gateways only through a particular device, i.e., the IoT hub. This contributes to the security of our approach, since guests cannot access IoT gateways/devices directly, as well as to its deployability, since no modification to gateways is required. 

We implemented a proof of concept DID registry and a distributed PDP using Hyperledger Fabric. 
Fabric is a private, permissioned blockchain technology, i.e., a blockchain system where membership is controlled. Fabric involves no monetary cost, low computational complexity, and insignificant delay. All DID-related cryptographic operations require less than 25 ms, even if they are executed in a mobile device. Furthermore, all registry operations and access control decisions implemented in Fabric are executed in less than 50 ms, apart from the DID creation process that requires 2.500 ms. 

\section*{Acknowledgment}
This research was supported by the EU funded Horizon 2020 project SOFIE (Secure Open Federation for Internet Everywhere), under grant agreement No. 779984.

\bibliographystyle{IEEEtran}
\bibliography{2019-iotsp}

\end{document}